\documentclass[twoside,a4paper,12pt]{article}

\usepackage{array,float,graphicx,amssymb,fancybox}
\usepackage{amsmath,bm,color}
\usepackage{fancyhdr}
\renewcommand{\theequation}{\arabic{section}.\arabic{equation}}

\newcommand{\R}{{\mathbb R}}

\renewcommand{\d}{{\mathrm d}}
\newcommand{\e}{{\mathrm e}}
\renewcommand{\i}{{\mathrm i}}
\newcommand{\p}{{\partial}}

\newcommand{\dast  }{{\displaystyle\ast}}

\newcommand{\be}{\begin{equation}}
\newcommand{\ee}{\end{equation}}
\newcommand{\ba}{\begin{array}}
\newcommand{\ea}{\end{array}}
\newcommand{\bal}{\begin{align}}
\newcommand{\eal}{\end{align}}
\newcommand{\bqa}{\begin{eqnarray}}
\newcommand{\eqa}{\end{eqnarray}}

\newcommand{\x}{\bm{x}}

\newcommand{\G}{\bm{G}}
\renewcommand{\k}{\bm{k}}
\renewcommand{\u}{\bm{u}}
\renewcommand{\v}{\bm{v}}
\newcommand{\U}{\bm{U}}
\newcommand{\V}{\bm{V}}

\newcommand{\K}{\bm{K}}
\renewcommand{\R}{\bm{R}}
\renewcommand{\r}{\bm{r}}

\title{Energy flux and dissipation of inhomogeneous plane waves in hereditary viscoelasticity}
\author{N. H. Scott\thanks{Email: n.scott@uea.ac.uk},
School of Mathematics,	
University of East Anglia,\\
Norwich Research Park,
Norwich NR4 7TJ. UK.
}

\date{27 April 2020}

\begin{document} 
\maketitle

\thispagestyle{fancy} \lhead{\emph{Proceedings of the Royal Society A,}  {\bf 475}:20190478  (2019)\\ 
http://dx.doi.org/10.1098/rspa.2019.0478\\ 
Published 6 November 2019}
\cfoot{}

\begin{center} Dedicated to the memory of Peter Chadwick FRS \end{center}

\begin{abstract}
Inhomogeneous small-amplitude plane waves of (complex) frequency $\omega$ are propagated through a linear dissipative material which displays hereditary viscoelasticity.  The energy density, energy flux and dissipation are quadratic in the small quantities, namely, the displacement gradient, velocity and velocity gradient, each harmonic with frequency $\omega$, and so give rise to attenuated constant terms as well as to inhomogeneous plane waves of frequency $2\omega$.  The quadratic terms are usually removed by time averaging but we retain them here as they are of comparable magnitude with the time-averaged quantities of frequency $\omega$.  A new relationship is derived in hereditary viscoelasticity that connects the amplitudes of the terms of the energy density, energy flux and dissipation that have frequency $2\omega$.  It is shown that the complex group velocity is related to the amplitudes of the terms with frequency $2\omega$ rather than to the attenuated constant terms as it is for homogeneous waves in conservative materials.

\vspace{1mm}\noindent
\end{abstract}

\section{Introduction}
\label{sec:1} 
\setcounter{equation}{0}

For the most general form of plane wave in a linear continuous medium, the particle displacement field $\u(\x, t)$ takes the complex exponential form
\be\label{1.1x}
\u(\x, t)  = 
\left\{\U\exp \i (\K\cdot \x - \omega t)\right\}^{+} 
\ee
where  $\i = \sqrt{-1}$ and 
\[ \U = \U^+ + \i \U^-,\quad \omega=\omega^+ +\i \omega^-,\quad \K = \K^+ + \i \K^- \]
are, respectively,  the complex wave  amplitude vector, the complex frequency  and the complex wave vector, all of which are constant.  Throughout this paper, the superscripts $^+$ and~$^-$ refer to real and imaginary parts of a complex quantity.
The real variables $\x$ and $t$ denote position and time, respectively.    
If the planes of constant 
 phase  $\K^+\cdot \x = \mbox{const.}$ and the planes of constant amplitude $\K^-\cdot \x = \mbox{const.}$ are not parallel, then   (\ref{1.1x}) is said to  represent an inhomogeneous 
plane wave.

Inhomogeneous plane waves arise in many different areas of mechanics both for conservative and for dissipative media, for example, Rayleigh and Stoneley waves in elasticity, electromagnetic radiation in wave guides, and surface and body waves in viscous fluids, viscoelastic solids and thermoelastic solids.  Inhomogeneous plane waves are important because all solutions of a linear problem may be written as superpositions, as finite sums or integrals, of waves of the form (\ref{1.1x}).

We show that the energy density, energy flux and dissipation in hereditary viscoelasticity are connected by an energy-dissipation equation, see (\ref{3.7x}) below.  These quantities are quadratic in the small quantities and so each consists of two parts, one an attenuated harmonic term with frequency $2\omega$ and the other an attenuated constant.  Usually, we time-average the governing equations by integrating over a cycle of the harmonic terms so that only the attenuated constant terms remain.  We find that these terms satisfy the relationship  (\ref{5.17x}) below in hereditary viscoelasticity.
Here, however, we explore the consequences of retaining the terms of frequency $2\omega$ in the theory of hereditary viscoelasticity because these terms are of the same order of magnitude as the attenuated constant terms.  We shall see these terms satisfy the relationship  (\ref{5.8x}) below.
For inhomogeneous waves we find that the (complex) group velocity is related to the amplitudes of the quadratic terms rather than to those of the (attenuated) constant terms as it is for homogeneous waves in conservative media, see (\ref{6.9x}) below.  

\pagestyle{fancy}
\fancyhead{}
\fancyhead[RO,LE]{\thepage}
\fancyhead[LO]{Energy flux and dissipation in hereditary  viscoelasticity}
\fancyhead[RE]{N. H. Scott}

The theory of inhomogeneous plane waves propagating through  continuous media has been given a detailed exposition by Boulanger \& Hayes \cite{BH}.  The idea of retaining the quadratic terms in the energy density, energy flux and dissipation in the theories of thermoelasticity and Kelvin-Voigt viscoelasticity has been explored by Scott \cite{scott1, scott2}.
Boulanger \cite{boul} obtained some of our results in the particular case of incompressible isotropic viscoelastic fluids.
Energy density, group velocity and dissipation in homogeneous and inhomogeneous plane waves have been much studied previously in a variety of contexts.  For example, 
Chadwick et al. \cite{chad1}  and Borejko \cite{borejko}  consider homogeneous and inhomogeneous wave propagation in a constrained elastic body, e.g. an incompressible or inextensible body, and discuss energy propagation and group velocity.
Cerveny \& Psencik \cite{cerveny2}  discuss  time-averaged and time-dependent energy-related quantities in inhomogeneous harmonic waves in anisotropic viscoelastic  media, especially Kelvin-Voigt viscoelasticity.   
Declercq et al. \cite{declerq}  discuss the history and properties of ultrasonic inhomogeneous waves, including complex frequency and  bounded beams.
Deschamps \& Huet \cite{deschamps1}   consider complex surface waves associated with inhomogeneous skimming and Rayleigh waves in linear elastodynamics.
Rodrigues Ferreira \& Boulanger \cite{rod}  extend the theory of     damped inhomogeneous waves to the finite-amplitude case in a deformed Blatz-Ko material.
Vashishth \& Sukhija \cite{vash} extend the theory of inhomogeneous waves to the case of  porous piezo-thermoelastic solids. 

The paper is constructed as follows.  In Section \ref{sec:2} we write down the constitutive equations of viscoelasticity in integral form.  Then in Section \ref{sec:3}  we use the equations of motion and double integral forms for the strain energy and the dissipation to derive an equation of energy balance including dissipation effects.  In Section \ref{sec:4} we derive the propagation condition for inhomogeneous plane waves in hereditary viscoelasticity.  In Section \ref{sec:5} we recall that the energy density, energy flux and energy dissipation are all quadratic in the small quantities occurring in the linear theory, e.g. the displacement and velocity gradients, and obtain expressions for them, each containing a constant term and one harmonic with frequency $2\omega$ where the linear quantities are harmonic with frequency~$\omega$.  We obtain some results valid for all dissipative media and further results proved here only for viscoelastic media.   Section \ref{sec:6} uses the dispersion relation to obtain a relation between the complex group velocity and the complex energy velocity, which is new to hereditary viscoelasticity, and concludes with some examples, namely,   linear elasticity, the Newtonian viscous fluid and the Kelvin-Voigt viscoelastic solid.
The double integrals for the strain energy and dissipation are evaluated in the Appendix.

\section{Constitutive equations}
\label{sec:2} 
\setcounter{equation}{0}

The particle velocity $\v(\x,t)$ is given by $\v(\x,t) =\dot{\u}(\x,t)$ where $\u$ denotes particle displacement, as at (\ref{1.1x}), and the superposed dot denotes the time partial derivative.  The components of the infinitesimal strain tensor $\bm{e}$ and the infinitesimal rate-of-strain tensor $\bm{d} = \dot{\bm{e}}$ are given by
\be \label{2.1x}
e_{ij} = \tfrac12(u_{i,\,j}+u_{j,\,i} ),\quad  d_{ij}=\tfrac12 (v_{i,\,j}+v_{j,\,i}),
\ee
respectively.  The notation $(\;)_{,\, j}$ denotes the spatial partial derivative $\p(\;)/\p x_j$.

The constitutive equations of anisotropic linear hereditary viscoelasticity are
\be\label{2.2x}
t_{ij} = \int_{-\infty}^tG_{ijkl}(t-\tau)d_{kl}(\tau) \, \d\tau
\ee
for the Cauchy stress $\bm{t}$, see for example \cite[Eq. (2.27)]{gurt},    in which twice-occuring roman suffices are summed over.  On putting $s=t-\tau$ in (\ref{2.2x}) we see that
\be \label{2.3x}
t_{ij}=\int_0^\infty G_{ijkl}(s)d_{kl}(t-s)\,\d s.
\ee
We assume that the components $G_{ijkl}(s)$ vanish for $s<0$ and satisfy  the further  properties 
\be\label{2.4x}
 G_{ijkl}(s)\geq 0,\quad G^\prime_{ijkl}(s)\leq 0,\quad
 \lim_{s\to0} G_{ijkl}(s) =  c^{\textrm{inst}}_{ijkl},\quad
\lim_{s\to\infty} G_{ijkl}(s) =  c_{ijkl},
\ee
where prime denotes differentiation with respect to argument.  The constants $ c^{\textrm{inst}}_{ijkl}$ are the instantaneous (small $t$) elastic moduli and the constants $c_{ijkl}$ are the equilibrium (large $t$) elastic moduli.  If the $c_{ijkl}$ all vanish then the material is a viscoelastic fluid rather than a viscoelastic solid.

The tensor components $G_{ijkl}(s)$ have the symmetries
\be\label{2.5x}
G_{ijkl}(s)=G_{jikl}(s) = G_{ijlk}(s),\quad s\geq 0,
\ee
because of the symmetries of $\bm{t}$ and $\bm{d}$.  We shall need the further symmetry property
\be\label{2.6x}
G_{ijkl}(s)=G_{klij}(s),\quad s\geq 0.
\ee
If this symmetry is present in the elastic moduli of a purely elastic material it implies the existence of a strain energy function. Gurtin \& Herrera \cite{gurt1}  have shown that $c^{\textrm{inst}}_{ijkl}$ and $c_{ijkl}$ each satisfy the symmetry (\ref{2.6x}), that is, the short and long time  material behaviour in hereditary viscoelasticity both satisfy the elastic symmetries.  Day \cite{day} has gone further.  He shows that (\ref{2.6x}) is obeyed for $0<s<\infty$ if and only if a certain work integral, namely, 
\[ W(\bm{e}) = \int_{-\infty}^\infty \bm{t}(\tau)\cdot \bm{d}(\tau) \,\d\tau, \]
is invariant under time reversal; i.e. if $W(\bm{e}(-t)) = W(\bm{e}(t))$.  However, it cannot be claimed that (\ref{2.6x}) has been proved for all $s>0$ and so we shall simply assume it, in common with most authors.

\section{Equations of motion and energy balance}
\label{sec:3} 
\setcounter{equation}{0}

For  hereditary viscoelasticity, the linearised equations of motion in the absence of body force are
\be\label{3.1x}
t_{ij,\,j}=\rho \dot{v}_i\,, 
\ee
where the mass density  $\rho$ may be taken to be constant.  
 
On multiplying (\ref{3.1x}) by $v_i$ we see that
\be\label{3.2x}
\dot{k}+r_{i,\,i}+t_{ij}v_{i,\,j}=0,\ee
 in which  $v_i$ are the components of the particle velocity $\v=\dot{\u}$ and  
\be\label{3.3x}
k=\tfrac12\rho v_iv_i,\quad r_i=  - t_{ji}v_j,
\ee
denote the kinetic energy and the components of the energy flux vector $\r$, respectively. 

By modelling an isotropic viscoelastic material as consisting of springs and dashpots connected in series and parallel Bland \cite{bland} and Hunter \cite{hunter} obtained expressions for the strain energy and dissipation in such a material as certain double integrals.  When generalised to the anisotropic case these double integrals become,  for the strain energy,
\be\label{3.4x}
\begin{aligned}
w &= \frac12\int_{-\infty}^t \int_{-\infty}^t
  G_{ijkl}(2t-\tau_1-\tau_2)d_{ij}(\tau_1)d_{kl}(\tau_2) \, \d\tau_1\d\tau_2, \\
   &=  \frac12\int^{\infty}_0\int^{\infty}_0
  G_{ijkl}(s_1+s_2)d_{ij}(t-s_1)d_{kl}(t-s_2)\,\d s_1\d s_2 
   \end {aligned}
   \ee
and for the dissipation
\be\label{3.5x}
\begin{aligned}
d &= - \int_{-\infty}^t \int_{-\infty}^t
  G^{\,\prime}_{ijkl}(2t-\tau_1-\tau_2)d_{ij}(\tau_1)d_{kl}(\tau_2) \, \d\tau_1\d\tau_2, \\
   &=  - \int^{\infty}_0\int^{\infty}_0
  G^{\, \prime}_{ijkl}(s_1+s_2)d_{ij}(t-s_1)d_{kl}(t-s_2)\,\d s_1\d s_2 , 
   \end {aligned}
   \ee
where, as before, prime denotes differentiation with respect to argument.
The expression (\ref{3.4x})$_1$ for the stored energy is attributed by Del Piero \& Desiri \cite{delpiero1} to Staverman \& Schwarzl \cite{staver}.  The determination of a suitable  expression for $w$ is discussed further by Golden \cite{golden1} and the references therein.
Buchen \cite{buchen} applied this model to plane waves in linear isotropic viscoelastic solids and Boulanger \cite{boul} applied it to plane waves in incompressible isotropic viscoelastic fluids.

Differentiating (\ref{3.4x})$_1$ under the integral sign with respect to $t$ gives
\[
\begin{aligned}
\dot{w} 
& = \frac12\int_0^t G_{ijkl}(t - \tau_2)d_{ij}(t) d_{kl}(\tau_2)\,\d\tau_2
+ \frac12\int_0^t G_{ijkl}(t - \tau_1)d_{ij}(\tau_1) d_{kl}(t)\,\d\tau_1 \\
 &\quad + \int_{-\infty}^t \int_{-\infty}^t
  G^{\,\prime}_{ijkl}(2t-\tau_1-\tau_2)d_{ij}(\tau_1)d_{kl}(\tau_2) \, \d\tau_1\d\tau_2.
\end{aligned}
\]
From (\ref{3.5x})$_1$ we see that the double integral above is equal to $-d$.
 From the symmetry property (\ref{2.6x}) the first two integrals above may be combined to give
 \[  \int_{-\infty}^tG_{ijkl}(t-\tau)d_{ij}(t)d_{kl}(\tau) \, \d\tau, \]
which from (\ref{2.2x}) is equal to $t_{ij}v_{i,\,j}$.
Combining these results leads to 
\be\label{3.6x}
\dot{w}=t_{ij}v_{i,\,j} -d.
\ee
Eliminating $t_{ij}v_{i,\,j}$ between this equation and (\ref{3.2x}) gives the energy balance equation in the form 
\be\label{3.7x}
\dot{e}+r_{i,\,i}= -d
\ee
with the total energy $e$ given by
\be\label{3.8x}
e=k+w.
\ee

\section{Inhomogeneous plane waves and the propagation condition}
\label{sec:4} 
\setcounter{equation}{0}

We now assume the small disturbance in the viscoelastic body to have the most general complex exponential plane wave form possible, so that the particle velocity $\v$  takes the form
\be\label{4.1x}
\v = \{\V\e^{\i\chi}\}^+
\ee
in which, as before, $^+$ denotes the real part of the quantity in braces and the phase factor $\chi$ is defined by
\be\label{4.2x}
\chi = \K\cdot \x - \omega  t  = \chi^++\i\chi^-,
\ee
where
\be\label{4.3x}
\chi^+  = \K^+\cdot \x - \omega^+  t\;,\quad \chi^-  = \K^-\cdot \x - \omega^-  t
\ee
are the real and imaginary parts of $\chi$ expressed in terms of those of $\K$ and $\omega$.  From   (\ref{4.3x}), we see that the wave amplitudes   (\ref{4.1x}) may be written
\be\label{4.4x}
\v = \{\V\e^{\i\chi^+}\}^+ \e^{-\chi^-}
\ee
from which it is clear that these wave amplitudes represent a sinusoidal travelling wave of frequency $\omega^+$ and wave vector $\K^+$ which is attenuated by the real exponential factor~$\e^{-\chi^-}$.

From the component form of the particle velocity  (\ref{4.1x}) and its definition  (\ref{2.1x}) we see that the symmetrised velocity gradient may be written
\be\label{4.5x}
d_{ij} = \left\{\tfrac12\i(V_iK_j+K_iV_j)\e^{\i\chi}\right\}^+
\ee
so that from its definition (\ref{2.3x}) the stress becomes
\be\label{4.6x}
t_{ij} = \left\{\i H_{ijkl}(\omega)V_kK_l \e^{\i\chi}\right\}^+
\ee
with
\be\label{4.7x}
H_{ijkl}(\omega) = \int_0^\infty G_{ijkl}(s)\e^{\i\omega s} \d s
\ee
denoting the half-range Fourier transform.

Applying the momentum balance equation (\ref{3.1x}) to the stress (\ref{4.6x}) and the velocity (\ref{4.1x}) leads to the propagation condition
\be\label{4.8x}
\left\{\rho\omega \delta_{ik}+\i H_{ijkl}(\omega) K_j
K_l\right\}V_k=0
\ee
with $\delta_{ik}$ denoting the components of the Kronecker delta.

\section{Energy density, energy flux and dissipation}
\label{sec:5} 
\setcounter{equation}{0}

\subsubsection*{General results}
We see from  (\ref{3.3x})--(\ref{3.5x}) and (\ref{3.8x}) that the energy density, energy flux and energy dissipation in the energy balance equation  (\ref{3.7x}) are quadratic in the small quantities and so, for inhomogeneous plane waves, are expressible as linear combinations of products of the form
\be\label{5.1x}
f(\bm{x}, t) =   \{A\e^{\i\chi}\}^+  \{B\e^{\i\chi}\}^+
\ee
in which $\chi$ continues to be given by   (\ref{4.2x})$_2$ and $A$ and $B$ are complex constants.  Using   (\ref{4.3x}), we may evaluate the product   (\ref{5.1x}) to obtain
\begin{align}
f(\bm{x}, t)& =   \{F\e^{2\i\chi}\}^+  + \overline{f}\e^{-2\chi^-}\label{5.2x}\\
\intertext{and from   (\ref{4.4x})}
f(\bm{x}, t)& =    \{F\e^{2\i\chi^+}\}^+\e^{-2\chi^-}  + \overline{f}\e^{-2\chi^-}\label{5.3x}
\end{align}
where
\be\label{5.4x}
F=\frac12AB\;,\quad \overline{f} = \left\{\frac12AB^\dast  \right\}^+
\ee
in which $F$ is a (usually) complex constant and $\overline{f}$ is a real constant.  Here and throughout, $^\dast  $ denotes the complex conjugate.  The first term of    (\ref{5.2x}) represents an inhomogeneous plane wave with phase factor $2\chi$, while the second consists of the real constant $\overline{f}$ attenuated by the real exponential factor $\e^{-2\chi^-}$.  From    (\ref{5.3x}) and (\ref{4.4x}), we see that the inhomogeneous plane wave is attenuated by the same factor and is sinusoidal with frequency $2\omega^+$.

To interpret $\overline{f}$, we follow \cite[Section 11.5]{BH} and integrate   (\ref{5.3x}) over a cycle of $\chi^+$ at constant $\chi^-$ to show that the mean value of $f$ is $ \overline{f}\e^{-2\chi^-}$, which depends on $\bm{x}$ and $t$ through $\chi^-$.  The real constant $\overline{f}$ is then regarded as a weighted mean of $f(\bm{x}, t)$.

We have already observed that the energy density, energy flux, and energy dissipation occurring in   (\ref{3.7x}) may be expressed as linear combinations of products of the form of   (\ref{5.1x}) and so, using   (\ref{5.2x}) and (\ref{5.3x}), we obtain
\begin{align}
e  &=  \{E\e^{2\i\chi^+}\}^+\e^{-2\chi^-}  + \overline{e}\,\e^{-2\chi^-} 
\label{5.5x}  \\
r_j  &=  \{R_j\e^{2\i\chi^+}\}^+\e^{-2\chi^-}   + \overline{r}_j\,\e^{-2\chi^-} 
\label{5.6x}  \\
d  &=  \{D\e^{2\i\chi^+}\}^+\e^{-2\chi^-}   + \overline{d}\,\e^{-2\chi^-} 
\label{5.7x}
\end{align}
in which $E, R_j, D$ are (usually) complex constants and $\overline{e},  \overline{r}_j,  \overline{d}$ are real constants.  These latter constants are the weighted means of $e, r_j, d$ as discussed above.

Previously, discussion of energy and dissipation has focused on the weighted means at the expense of terms involving the complex quantities $E, R_j, D$, see for example \cite{BH}, often on the grounds that these terms do not contribute when averaged over a cycle of $\chi^+$.  However, the energy-dissipation equation  (\ref{3.7x}) is valid for all $\bm{x}$ and $t$, without averaging, and the neglected terms are of the same order of magnitude (before averaging) as the retained terms. 

 It is our chief purpose here to explore the consequences of retaining the attenuated harmonic terms on an equal footing with the weighted means.

On substituting   (\ref{5.5x})--(\ref{5.7x}) into  (\ref{3.7x}) and equating the coefficients of the attenuated harmonic terms, and those of the purely attenuated terms, we obtain
\begin{align}
& \omega E - \K\cdot \R + \i D =0  
\label{5.8x}\\[1mm]
& \omega^-\,\overline{e} - \K^-\cdot \overline{\r} + \overline{d} =0  \label{5.9x}
\end{align}
respectively.  Equation (\ref{5.8x}) has appeared at \cite[Eq. (44)]{scott1} and \cite[Eq. (4.15)]{scott2}.   Equation (\ref{5.9x}) has appeared at \cite[Eq. (76)]{scott3}, \cite[Eq. (45)]{scott1},  \cite[Eq. (4.17)$_2$]{scott2}  as well as at  \cite[Eq. (11.5.8)]{BH}, where it was derived by a different method.  Equations (\ref{5.8x}) and (\ref{5.9x}) have a wide range of validity, not only in viscoelasticity, since they are valid for any system that has an energy-dissipation equation of the form  (\ref{3.7x}) with energy density, energy flux, and energy dissipation being quadratic in the small quantities and taking the forms given by   (\ref{5.5x})--(\ref{5.7x}).

We may conclude from    (\ref{5.9x}) that if both $\omega$ and $\K$ are real, then $\overline{d}=0$ and there is no weighted mean dissipation.  Alternatively, if there is dissipation ($\overline{d}\neq 0$),  then we may draw the conclusion from    (\ref{5.9x}) that not both of $\omega$ and $\K$ can be real.

\subsubsection*{Hereditary viscoelasticity}

For the kinetic energy $k$ defined by (\ref{3.3x})$_1$ we take $\v$ in the form (\ref{4.4x}) and use the formula (\ref{5.3x}) to show that
\[ k= 
\frac{\rho}{4}\left\{\V\cdot\V\e^{2\i\chi^+}\right\}^+\e^{-2\chi^-}
+ \frac{\rho}{4}\V\cdot\V^\dast\e^{-2\chi^-}.
\]
By adding this equation to (\ref{A7}) we obtain the equation  (\ref{5.5x}) for the total energy $e=k+w$ defined at (\ref{3.8x}) with $E$ and $\overline{e}$ given by  (\ref{5.10x}) and  (\ref{5.13x}), respectively.

For hereditary viscoelasticity, we may obtain explicit expressions for the quantities $E, R_j, D$ and  $\overline{e},  \overline{r}_j,  \overline{d}$ occurring in    (\ref{5.5x})--(\ref{5.7x})   by substituting the inhomogeneous plane wave forms of  (\ref{4.1x}) into   (\ref{3.3x})--(\ref{3.5x}) and (\ref{3.8x}) for the energy density, energy flux, and energy dissipation and using  (\ref{5.2x}) and (\ref{5.4x}):
\allowdisplaybreaks
\begin{align}
E     &= 
  \frac{\rho}{4}V_iV_i + \frac{\i}{4} H_{ijkl}^\prime (\omega)V_iK_jV_kK_l 
\label{5.10x}\\
R_j &= -\frac{\i}{2} H_{ijkl} (\omega) V_iV_kK_l 
\label{5.11x}\\
 D   &= -\frac{1}{2}(H_{ijkl} (\omega) +\omega H^\prime_{ijkl} (\omega))V_iK_jV_kK_l 
 \label{5.12x}\\
\overline{e} &= 
\frac{\rho}{4}V_iV^\dast_i 
+ \frac{1}{4\omega^+} H^-_{ijkl} (\omega) V_i^\dast K_j^\dast V_kK_l
\label{5.13x}\\
\overline{r}_j&= \left\{- \frac{\i}{2} H_{ijkl} (\omega) V^\dast_iV_kK_l \right\}^+
\label{5.14x}\\
\overline{d} &= -\frac{1}{2}\left(\frac{\omega^-}{\omega^+}H^-_{ijkl} (\omega) -H^+_{ijkl} (\omega) \right)
V_iK_jV^\dast_kK_l^\dast.
\label{5.15x}
\end {align}
We may use the symmetry property (\ref{2.6x}), in the form 
\be\label{5.16x}
H_{ijkl}(\omega) = H_{klij}(\omega),
\ee
to verify that $\overline{e}$ is real.
The reality of $\overline{r}_j$ is clear.   It also follows from (\ref{2.6x})  that $\overline{d}$ is real. 
As in the general case, the non-vanishing of $\overline{d}$ implies that not both of $\omega$ and $\K$ can be real.

We may verify the general equation (\ref{5.8x}) in the present case of hereditary viscoelasticity by substituting for $E, \R$, and $D$ from    (\ref{5.10x})--(\ref{5.12x}) into (\ref{5.8x}) and observing that it is satisfied. 
 In the same way, we may use   (\ref{5.13x})--(\ref{5.15x}) for the weighted means $\overline{e},  \overline{r}_j,  \overline{d}$ to derive the identity
\be\label{5.17x}
 \omega \overline{e} - \K\cdot \overline{\r} + \i \overline{d} = 0.
\ee
Equations (\ref{5.8x}) and (\ref{5.17x}) have the same form, the first involving the amplitudes of the attenuated harmonic terms and the second involving the weighted means, but it should be remembered that  (\ref{5.8x}) has general validity while  (\ref{5.17x}) has been demonstrated here only for viscoelasticity.

Bearing in mind that $\overline{e},  \overline{r}_j,  \overline{d}$ are real, we may take real and imaginary parts of  (\ref{5.17x}) to obtain
\be\label{5.18x}
\overline{\r}\cdot\K^+ = \omega^+ \,\overline{e}\;,\quad \overline{\r}\cdot\K^- =  \omega^- \,\overline{e} +\overline{d}.
\ee
As might be expected from the absence of $\overline{d}$,  (\ref{5.18x})$_1$ is valid also for conservative media and was proved by Hayes \cite[Eq. (4.8)$_2$]{hayes2}.  Equation (\ref{5.18x})$_2$ simply verifies for  hereditary viscoelasticity the general result   (\ref{5.9x}).

\section{Dispersion relation, group velocity and energy velocity}
\label{sec:6} 
\setcounter{equation}{0}

In theories of continuous media, one typically derives from the propagation conditions, such as (\ref{4.8x}), an equation giving the frequency as a function of the wave vector:
\be\label{6.1x}
\omega=\omega(\K)
\ee
known as the dispersion relation.  In the present case of hereditary viscoelasticity we could obtain from (\ref{4.8x}) the dispersion relation  (\ref{6.1x}) in the implicit form
\[ \det \left\{\rho\omega \delta_{ik}+\i H_{ijkl}(\omega) K_j
K_l\right\}=0. \] 
However, it proves more convenient to consider the propagation condition (\ref{4.8x}) in its original form
\be\label{6.2x}
\rho\omega V_i+\i H_{ijkl}(\omega) K_jV_kK_l=0. 
\ee
 
It follows from  (\ref{6.1x}) and (\ref{6.2x}) that $\omega$ and $\V$  depend on $\K$ but not on its complex conjugate    $\K^\dast $.  Then  $E, R_j, D$, defined by  (\ref{5.10x})--(\ref{5.12x}), are functions of $\K$ but not   $\K^\dast $.  Clearly, $\omega^\dast$ and $\V^\dast$ are functions of $\K^\dast $ but not $\K$.  It follows that the real quantities $\overline{e},  \overline{r}_j,  \overline{d}$, defined by   (\ref{5.13x})--(\ref{5.15x}), are functions of both $\K$ and  $\K^\dast $. 

Now  (\ref{5.8x}) holds for all possible complex wave vectors $\K$ and so may be regarded as an identity in $\K$.  Allowing the operator $\p/\p K_p$ to act upon this equation then gives
\be\label{6.3x}
\frac{\p\omega}{\p K_p}E-R_p = - \left\{\omega\frac{\p E}{\p K_p} - K_j\frac{\p R_j}{\p K_p} +\i \frac{\p D}{\p K_p}\right\}.
\ee
In the general case of a linear dissipative material, we do not have explicit expressions for  $E, R_j, D$ and so can make no further progress.

Equation (\ref{5.9x}) also holds for all possible complex wave vectors $\K$, but since its terms depend explicitly also on $\K^\dast $, it is to be regarded as an identity in each of the six quantities $K_p^+, K_p^-, p=1,2,3$.
Equivalently,  (\ref{5.9x}) is an identity in each of the six components of $\K$ and $\K^\dast $, with $\K^\dast $ now regarded as independent of $\K$.  Therefore, we rewrite  (\ref{5.9x}) as 
\[ (\omega-\omega^\dast )\,\overline{e}-(K_j-K_j^\dast )\,\overline{r}_j+2\i \overline{d}=0 \]
and allow $\p/\p K_p$ to act upon it, bearing in mind that $\omega$ depends only on $\K$ and $\omega^\dast $ depends only on $\K^\dast $, to obtain
\be\label{6.4x}
\frac{\p \omega}{\p K_p}\overline{e} - \overline{r}_p = -2\i\left\{\omega^-\frac{\p\, \overline{e}}{\p K_p} - K_j^- \frac{\p\, \overline{r}_j}{\p K_p} + \frac{\p\, \overline{d}}{\p K_p} \right\}.
\ee
As with  (\ref{6.3x}), we do not have explicit expressions for $\overline{e},  \overline{r}_j,  \overline{d}$ in the general case of a linear dissipative material and so can make little further progress.

There is, however, one deduction we can make from  (\ref{6.4x}).  In the case of homogeneous waves in a dissipationless system,  the complex wave vector $\K$ is replaced by the real one $\k$, the frequency $\omega$ also is real, and the dissipation $d$ vanishes, so that
\[ \omega^-=0\;,\quad K_j^-=0\;,\quad \overline{d}=0 \]
and  (\ref{6.4x}) reduces to 
\be\label{6.5x}
\overline{r}_p=  \overline{e} \,\frac{\p\, \omega}{\p K_p} 
\ee
valid for homogeneous waves in a general dissipationless system as proved by Hayes \cite[Eq. (20)]{hayes1}.  

We return to our discussion of energy and dissipation in hereditary viscoelasticity and seek a connection between the complex group velocity $\p \omega/\p \K$ and quantities $E$ and $\R$ already defined at (\ref{5.10x}) and (\ref{5.11x}).   We apply the operator $\p/\p K_p$ to  (\ref{6.2x}) and contract the resulting equation with $V_i$ to obtain
\begin{align}
 & \left\{ \rho\omega V_i + \i H_{klij}K_jV_kK_l  \right\}\frac{\p V_i}{\p K_p} \nonumber \\
 &\quad + \left\{  \rho V_iV_i+\i H^\prime_{ijkl}V_iK_jV_kK_l \right\}\frac{\p\omega}{\p K_p} \label{6.6x}  \\
  &\quad\quad\; + \i H_{ipkl}V_iV_kK_l + \i H_{ijkp}V_iK_jV_k =0.
  \nonumber
\end{align}
We may use the symmetry property (\ref{5.16x}) 
to show from the propagation condition (\ref{6.2x}) that the first term of (\ref{6.6x}) vanishes.  Furthermore, we contract  (\ref{6.2x}) with $V_i$ and use the result to eliminate $\rho V_iV_i$ from the second term of  (\ref{6.6x}).  We then use the definitions (\ref{5.10x}) and (\ref{5.11x}) to show that the remaining two terms of (\ref{6.6x}) reduce to
\be\label{6.7x}
R_p = E\frac{\p\omega}{\p K_p}.
\ee
This is an important result in the theory of inhomogeneous waves in dissipative media, here demonstrated for hereditary viscoelasticity. It has previously been demonstrated for thermoelasticity, see Scott \cite[Eq. (69)]{scott1}, and for Kelvin-Voigt viscoelasticity, see Scott \cite[Eq. (5.1)]{scott2}.   One might expect this result to have a wider validity in the theory of dissipative media, but this has not been demonstrated.  

We define a complex energy velocity $\G$ associated with the attenuated harmonic terms of the energy-dissipation equation by
\be\label{6.8x}
\G=\frac{\mbox{attenuated harmonic energy flux}}{\mbox{attenuated harmonic energy density}} = 
\frac{\R}{E}
\ee
provided $E\neq 0$, which from  (\ref{6.7x}) may be written
\be\label{6.9x}
\G= \frac{\p\omega}{\p \K}.
\ee
In terms of $\G$,  (\ref{5.8x}) becomes
\be\label{6.10x}
\G\cdot\K = \omega + \i D/E.
\ee

The energy velocity vector more usually considered in the literature, that associated with the weighted mean quantities, is defined by
\be\label{6.11x}
\bm{g}=\frac{\mbox{weighted mean energy flux}}{\mbox{weighted mean energy density}}
= \frac{\overline{\bm{r}}}{\;\overline{e}\;}
\ee
a purely real vector. 
In terms of $\bm{g}$,  (\ref{5.17x}) becomes
\be\label{6.12x}
\bm{g\cdot K} = \omega + \i \overline{d}/\overline{e}
\ee
comparable with   (\ref{6.10x}), and has real and imaginary parts
\be\label{6.13x}
\bm{g\cdot K}^+ = \omega^+,\quad \bm{g\cdot K}^- = \omega^- + \overline{d}/\overline{e}.
\ee

\subsubsection*{Examples}
\paragraph{Linear elasticity}  $G_{ijkl}(s)= c_{ijkl}h(s)$, where $h(s)$ is the Heaviside step function: 
$h(s)=0\mbox{ if }s<0\mbox{ and } h(s)=1\mbox{ if }s\geq0.$
The quantities $c_{ijkl}$ are the usual elastic moduli.  From (\ref{4.7x}) we see that $H_{ijkl}=  \left(-1/\i\omega \right) c_{ijkl}$ so that the propagation condition (\ref{4.8x})  becomes 
\be\label{6.14x}
\left\{\rho\omega^2 \delta_{ik} - c_{ijkl} K_j
K_l\right\}V_k=0
\implies
\det \left\{\rho\omega^2 \delta_{ik} - c_{ijkl} K_j
K_l\right\}=0. 
\ee
Thus we see that in the dispersion relation (\ref{6.1x}), $\omega$ is homogeneous of degree one in $\K$. Elasticity is a conservative theory, so that there is no dissipation,  and the counterparts in elasticity of the present equations  (\ref{6.10x}) and  (\ref{6.12x}), with $D=\overline{d}=0$, are simple consequences of the fact that $\omega$ is homogeneous of degree one in $\K$.  
Many of the present results have been obtained for constrained elastic materials by Chadwick et al. \cite[Eqns (4.16), (4.22) and (4.24)]{chad1}, see also Borejko \cite[Eqns (3.19), (4.18), (4.20), (4.24), (4.27) and (4.30)]{borejko}.

\paragraph{Newtonian viscous fluid}$G_{ijkl}(s)= \eta_{ijkl}\delta(s)$, where $\delta(s)$ is the Dirac delta function: 
$ \delta(s)=0\mbox{ if } s\neq0 \mbox{ and } \int_0^\infty f(s)\delta(s) \,\d s = f(0)$ for any function $f$ continuous at $x=0$.  For the Newtonian viscous fluid we have  
\be\label{6.15x}
 \eta_{ijkl} =  (\kappa-\tfrac23\mu) \delta_{ij}\delta_{kl} + 
\mu(\delta_{ik}\delta_{jl}+\delta_{il}\delta_{jk}),
\ee
 where $\kappa$ is the bulk viscosity and $\mu$ the shear viscosity. Thus, from (\ref{4.7x}) we see that for viscous fluids 
 $H_{ijkl}=\eta_{ijkl}$,
 independent of $\omega$.  The propagation condition (\ref{4.8x}) then becomes 
 \[ \left\{\rho\,\i\omega \delta_{ik} - \eta_{ijkl} K_j
K_l\right\}V_k=0,
\]
where $\eta_{ijkl}$ is given at  (\ref{6.15x}).  This is discussed further in \cite{scott3} in which the present equations  (\ref{5.9x}) and (\ref{5.17x})   are derived for compressible viscous fluids.

\paragraph{Kelvin-Voigt viscoelasticity}  $G_{ijkl}(s)= c_{ijkl}h(s) + \eta_{ijkl}\delta(s)$, which is simply a linear combination of linear elasticity and  the Newtonian viscous fluid described above.  Then 
\[ H_{ijkl}(\omega) = - \frac{1}{\i\omega} c_{ijkl} + \eta_{ijkl}
\]
so that the propagation condition (\ref{4.8x}) becomes
\[ \left\{\rho\omega^2 \delta_{ik} - (c_{ijkl} - \i\omega\eta_{ijkl}) K_jK_l\right\}V_k=0,
  \]
the same as \cite[Eq. (3.9)]{scott2} derived directly for a Kelvin-Voigt viscoelastic material.  The present equations  (\ref{5.9x}) and (\ref{5.17x}), and many others,   are derived directly for a Kelvin-Voigt viscoelastic material in \cite{scott2}.\\






\noindent
\emph{Acknowledgement}  {The author is much indebted to the referees for helpful suggestions and supplying a new reference.}



\vskip2pc



\appendix
\renewcommand{\theequation}{\Alph{section}\arabic{equation}}

\section{Appendix}
\setcounter{equation}{0}

\subsection*{Evaluation of the integral $(\ref{3.4x})_2$ for $w$}

From (\ref{4.5x}) and (\ref{5.3x}) we see that
\be\label{A1}
\begin{aligned}
d_{ij}(t-s_1)d_{kl}(t-s_2) 
&= -\frac18\left\{ (V_iK_j+K_iV_j)(V_kK_l+K_kV_l)\e^{2\i\chi^+-2\chi^-}\e^{\i\omega(s_1+s_2)}  \right\}^+  \\
&\quad + \frac18\left\{ (V_iK_j+K_iV_j)(V^\dast_kK^\dast_l+K^\dast_kV^\dast_l)  \e^{-2\chi^-} \e^{\i\omega s_1-\i\omega^\dast s_2}  \right\}^+.
\end{aligned}
\ee
We can show from this equation that the integrand of (\ref{3.4x})$_2$ may be written
\be\label{A2}
\begin{aligned}
G_{ijkl}(s_1+s_2)d_{ij}(t-s_1)&d_{kl}(t-s_2) 
= -\frac12\left\{ G_{ijkl}(s_1+s_2)V_iK_jV_kK_l\e^{2\i\chi^+-2\chi^-}\e^{\i\omega(s_1+s_2)}  \right\}^+  \\
& + \frac12 G_{ijkl}(s_1+s_2)V_iK_jV^\dast_kK
^\dast_l  \e^{-2\chi^-} \e^{ -\omega^- (s_1+s_2)}
\cos \omega^+(s_1-s_2),
\end{aligned}
\ee
where (\ref{2.5x}) has been used.  By taking the complex conjugate of 
\(  G_{ijkl}(s_1+s_s)V_iK_jV^\dast_kK
^\dast_l \) 
and using the symmetry (\ref{2.6x}) we see that this quantity is real leading to the last line of~(\ref{A2}).

Dropping temporarily the suffixes $_{ijkl}$, we evaluate the integrals
\begin{align}
I_1 
&= \int_0^\infty\int_0^\infty G(s_1+s_2) \e^{\i\omega(s_1+s_2)}
\, \d s_1 \d s_2,  
\label{A3}  \\
I_2
&= \int_0^\infty\int_0^\infty G(s_1+s_2) \e^{-\omega^-(s_1+s_2)} \cos\omega^+(s_1-s_2)\, \d s_1 \d s_2,
\label{A4}
\end{align}
as these integrals will be needed in performing the double integral in (\ref{3.4x})$_2$ with integrand (\ref{A2}).
By means of the substitutions $\alpha = s_1+s_2$ and $\beta = -s_1+s_2$ we see that
\begin{align}
I_1
&=\frac12\int_0^\infty\int_{-\alpha}^\alpha G(\alpha)\e^{\i\omega\alpha}\,\d\beta\,\d\alpha  \nonumber \\
\intertext{and on performing the $\beta$ integral}
I_1
&= \int_0^\infty G(\alpha)\alpha\, \e^{\i\omega\alpha}\,\d\alpha
  = \frac{1}{\i}\frac{\d}{\d\omega} \int_0^\infty G(\alpha) 
 \e^{\i\omega\alpha}\,\d\alpha \nonumber\\
&= -\i H^\prime(\omega),
\label{A5}
\end{align}
where  $H(\omega) = \int_0^\infty G(\alpha) 
\e^{\i\omega\alpha} \d\alpha$, see (\ref{4.7x}).
Also, we see that
\begin{align}
I_2 
&= \frac12\int_0^\infty\int_{-\alpha}^\alpha G(\alpha) \e^{-\omega^-\alpha} \cos(\omega^+\beta)\,\d\alpha\,\d\beta
\nonumber \\
& = \frac{1}{\omega^+} \int_0^\infty G(\alpha) \e^{-\omega^-\alpha}\, \sin(\omega^+\alpha)\,\d\alpha 
\nonumber \\
&= \frac{1}{2\i\omega^+}\int_0^\infty G(\alpha) 
\left(\e^{\i\omega\alpha} - \e^{(\i\omega)^\dast \alpha}\right)\d\alpha
\nonumber \\
&=  \frac{1}{2\i\omega^+}\left( H(\omega) - \left\{H(\omega)\right\}^\dast \right)
\nonumber \\
&= \frac{H^-(\omega)}{\omega^+},
\label{A6}
\end{align}
where $H^-(\omega)$ denotes $\left\{H(\omega)\right\}^-$,  the imaginary part of $H(\omega)$.

We  now evaluate $w$ by performing the integrals in (\ref{3.4x})$_2$ taking the integrand in the form (\ref{A2}).  The ensuing integrals $I_1$ and $I_2$ are given by (\ref{A5}) and (\ref{A6}), respectively, so that we finally obtain
\be\label{A7}
 w  =  \{W\e^{2\i\chi^+}\}^+\e^{-2\chi^-}  + \overline{w}\,\e^{-2\chi^-},
\ee
where
\[ W = \frac{\i}{4}H^\prime_{ijkl}(\omega)V_iK_jV_kK_l,
\quad \overline{w} = \frac{1}{4\omega^+}
 H^-_{ijkl}(\omega)V_iK_jV^\dast_kK^\dast_l.
\]

\subsection*{Evaluation of the integral $(\ref{3.5x})_2$ for $d$}

To evaluate $d$ by means of (\ref{3.5x})$_2$ we need to evaluate the same double integral as in (\ref{3.4x})$_2$ except that $G_{ijkl}(s_1+s_2)$ is replaced by its derivative $G^{\,\prime}_{ijkl}(s_1+s_2)$.  Therefore the integrand of (\ref{3.5x})$_2$ must be replaced by (\ref{A2}) except that  $G_{ijkl}(s_1+s_2)$ is replaced by its derivative $G^{\,\prime}_{ijkl}(s_1+s_2)$.  In place of $I_1$ and $I_2$ above we must therefore evaluate the integrals
\begin{align}
I_3 
&= \int_0^\infty\int_0^\infty G^{\,\prime}(s_1+s_2) \e^{\i\omega(s_1+s_2)}
\, \d s_1 \d s_2 
\label{A8}  \\
I_4
&= \int_0^\infty\int_0^\infty G^{\,\prime}(s_1+s_2) \e^{-\omega^-(s_1+s_2)} \cos\omega^+(s_1-s_2)\, \d s_1 \d s_2.
\label{A9}
\end{align}
Substituting $\alpha = s_1+s_2$ and $\beta = -s_1+s_2$ as before, we see that
\[
I_3
 =\frac12\int_0^\infty\int_{-\alpha}^\alpha G^{\,\prime}(\alpha)\e^{\i\omega\alpha}\,\d\beta\,\d\alpha  \nonumber \\
 = \int_0^\infty G^{\,\prime}(\alpha)\alpha\, \e^{\i\omega\alpha}\,\d\alpha.  
 \]
Integrating by parts gives
\begin{align}
I_3 &= \left[  G (\alpha)\alpha\, \e^{\i\omega\alpha} \right]_0^\infty  - \int_0^\infty G(\alpha)\left(\e^{\i\omega\alpha} + \i\omega\alpha \e^{\i\omega\alpha}\right) \d\alpha \nonumber \\  
&= - H(\omega) - \omega H^\prime(\omega),
\label{A10}
\end{align}
where  the integrated out limits vanish and the integral (\ref{A5}) has been used.

\allowdisplaybreaks
Also, we see that
\begin{align}
I_4 
&= \frac12\int_0^\infty\int_{-\alpha}^\alpha G^{\,\prime}(\alpha) \e^{-\omega^-\alpha} \cos(\omega^+\beta) \,\d\beta\,\d\alpha
\nonumber \\
&= \frac{1}{\omega^+} \int_0^\infty G^{\,\prime}(\alpha) \e^{-\omega^-\alpha}\, \sin(\omega^+\alpha)\,\d\alpha 
\nonumber \\
 &= \frac{1}{\omega^+}\left[G(\alpha)\e^{-\omega^-\alpha} \sin(\omega^+\alpha)\right]_{\alpha=0}^\infty  \nonumber \\
 &\qquad  - \frac{1}{\omega+}\int_0^\infty G(\alpha)\left\{ -\omega^-\e^{-\omega^-\alpha} \sin(\omega^+\alpha) + \omega^+\e^{-\omega^-\alpha} \cos(\omega^+\alpha)  \right\}\d\alpha
 \nonumber \\
  &= \frac{\omega^-}{\omega^+} \int_0^\infty G(\alpha) \e^{-\omega^-\alpha}\, \sin(\omega^+\alpha)\,\d\alpha   
  - \int_0^\infty G(\alpha) \e^{-\omega^-\alpha}\, \cos(\omega^+\alpha)\,\d\alpha   
  \nonumber \\
    &= \frac{\omega^-}{\omega^+} H^-(\omega) - \frac12\int_0^\infty G(\alpha)\left(\e^{\i\omega\alpha} + \e^{(\i\omega)^\dast\alpha}\right) \d\alpha  \nonumber \\
&=   \frac{\omega^-}{\omega^+} H^-(\omega)  - \frac12\left( H(\omega) + \left\{H(\omega)\right\}^\dast \right)
\nonumber \\
&= \frac{\omega^-}{\omega^+}H^-(\omega) - H^+(\omega).
\label{A11}
\end{align}
We  now evaluate $d$ by performing the integrals in (\ref{3.5x})$_2$ using the integrals $I_3$ and $I_4$ given by (\ref{A10}) and (\ref{A11}), respectively, so that we finally obtain equation (\ref{5.7x}) with $D$ and $\overline{d}$ given by (\ref{5.12x}) and (\ref{5.15x}), respectively.

\end{document}